\documentclass[a4paper,11pt,twocolumn]{article}
\usepackage[margin={1.5cm,3cm,1.5cm,0.1cm}]{geometry}
\usepackage[colorlinks,citecolor=blue]{hyperref}
\hypersetup{pdfauthor=Hamid Saleem \& Usman Saeed,pdftitle=Nonlinear Drift Waves and Density Cavities in Ionosphere,pdfsubject={Drift Waves},pdfcreationdate=\today}
\usepackage[utf8]{inputenc}
\usepackage{amsmath,amsthm,amssymb}
\usepackage[sorting=none,maxnames=4,minitems=2,bibstyle=numeric,citestyle=authoryear,maxcitenames=1]{biblatex}
\usepackage{graphicx}
\usepackage{dcolumn}
\usepackage{bm}
\usepackage{physics}
\usepackage[utf8]{inputenc}
\usepackage[T1]{fontenc}
\usepackage{mathptmx}
\usepackage{etoolbox}
\usepackage{authblk}

\begin{document}
	
	
\title{Nonlinear Drift Waves and Electrostatic Density Cavities in Upper Ionosphere}
\date{{}}	
\author[1,2]{H. Saleem\thanks{saleemhpk@hotmail.com}}
\author[1,3]{U. Saeed\thanks{usman.exe@gmail.com}}

\affil[1]{Theoretical Research Institute, Pakistan Academy of Sciences (TRIPAS), 3-Constitution Avenue, G-5/2, Islamabad (44000), Pakistan.}
\affil[2]{School of Natural Sciences (SNS), National University of Sciences and Technology (NUST), H-12, Islamabad (44000), Pakistan.}
\affil[3]{Theoretical Physics Department (TPD), National Centre for Physics, Quaid-i-Azam University Campus, Shahdra Valley Road, Islamabad}

	
	\maketitle
	
	\begin{abstract}
Nonlinear drift waves are investigated to explain the mechanism for the creation of one dimensional (1D) and two dimensional (2D) electrostatic density cavities in the presence of density gradient in direction perpendicular to the geomagnetic field. Ion motion along field lines has to be neglected for the formation of density depletion because nonlinear contribution of their parallel motion overcomes the effect of drift waves which create negative nonlinear potential. It is pointed out that a straightforward but appropriate model for drift waves can explain the generation of density cavities having widths in xy-plane if the external magnetic field is assumed to be along z-axis i.e. ${\bf B}_0 = B_0 \hat{z}$. The general nonlinear equation for drift waves takes the form of  Zakharov-Kuznetson (ZK) equation when ion nonlinear vorticity term is ignored and it admits the solution in the form of 2D solitary wave. In 1D case, the ZK equation reduces to Korteweg de Vries (KdV) equation which admits solution in the form of 1D soliton. In both cases, the nonlinear potential becomes negative. The order of magnitude of the amplitude and width of density cavities is found to be in agreement with the satellite observations in upper ionosphere.
\end{abstract}
	
	\vspace{0.5 cm}
	\section{Introduction}
The existence of solitary structures is almost ubiquitous in Earth's ionosphere and magnetosphere at different altitudes particularly within the auroral latitudes. The electrostatic solitary structures were observed in upper ionosphere (or lower magnetosphere) by Freja satellite \parencite{dovner1994freja, Wahlund1994a, Wahlund1994b} at altitude of about $1700$ $km$ as well as in magnetosphere at altitude of about $8000$ kilometers ($km$) by Viking satellite \parencite{bostrom1988characteristics}. The Freja satellite observed two types of density cavities; one is associated with lower hybrid waves known as lower hybrid cavitons \parencite{eriksson1994freja} and the other seems to be created in the presence of Boltzmann electrons \parencite{dovner1994freja} which indicates the low frequency wave activity in this region. The low frequency electrostatic waves such as ion acoustic wave (IAW) and drift wave (DW) exist in the limit of inertialess electrons. But nonlinear IAWs produce hump solitons \parencite{sagdeev1966reviews, washimi1966, yu1980fully} and cannot give rise to electrostatic structures with negative potential. Cairns et al. \parencite{Cairns1995} showed that the density dips can be created by nonlinear IAWs if electrons are assumed to follow the non-thermal Cairns distribution instead of Boltzmann distribution in ionosphere. They showed that the small size electrostatic structures of the order of $50 \lambda_{De}$ (where $\lambda_{De}$ is the electron Debye length) along magnetic field direction can be formed by nonlinear IAWs if electrons follow Cairns distribution. They considered the propagation of IAWs only along the ambient magnetic field, therefore the effects of magnetic field do not appear in this investigation. The IAWs have been investigated extensively in literature \parencite{steffy2017existence, saleem2020theoretical, lakhina2021electrostatic, varghese2022morphology}. Many authors also investigated drift wave like modes to investigate vortex like structures in upper ionosphere \parencite{shukla1995vortex, johnson1995nonlinear}.

A few years ago \parencite{Saleem2024}, it was shown that the pure drift waves create density dips in magnetized plasma because the nonlinear term in the Korteweg de-Vries (KdV) equation turns out to be negative if ion motion along the ambient magnetic field lines is ignored. In initial studies of nonlinear drift waves \parencite{nozaki1974propagation, gell1977drift}, the ion field aligned motion was taken into account and the ion fluid velocity was normalized with ion acoustic speed $c_s=(\frac{T_e}{m_i})^{1/2}$ in the derivation of KdV equation using reductive perturbation (RPM) method. If drift wave dynamics are considered including the ion parallel motion, then this wave couples with IAW and usually the ion velocity is normalized with $c_s$ in the derivation of solitary wave equations which is not the phase speed of drift waves. Therefore, the fundamental dynamics of nonlinear drift waves are not seen clearly if contribution of ion parallel motion is taken into account.

On the other hand, Hasegawa and Mima \parencite{hasegawa1978pseudo} investigated the two dimensional (2D) drift waves to explain the electrostatic turbulence observed in magnetized plasma. They derived a nonlinear equation for pure 2D drift waves neglecting the contribution of ion parallel motion. However, the Hasegawa-Mima (HM) equation is valid for short wavelength 2D drift waves in the limit $\rho_s k_{\perp} \simeq O (1)$ (where $\rho_s=\frac{c_s}{\Omega_i}$, $c_s=(\frac{T_e}{m_i})^{1/2}$ is ion acoustic speed, $\Omega_i=\frac{eB_0}{m_i c}$ is the ion gyro frequency, and $k_{\perp}$ is the component of wave vector perpendicular to external magnetic field ${\bf B_0}$). The HM equation also admits dipolar vortex solution when the wave amplitude is assumed to be slowly varying in the plane perpendicular to ${\bf B}_0$. If Magnetic field is assumed to be along z-axis i.e. ${\bf B}_0= B_0 \hat{z}$, then the electrostatic potential is assumed to be the function of $(x,y)$ coordinates. 

Recently \parencite{saleem2026exploring}, a general nonlinear equation for 2D   drift waves has been derived which reduces to HM equation in one limit and to Zakharov-Kuznetson (ZK) equation in another limit. It has been shown that the ion vorticity term becomes smaller than the electron nonlinear density term, particularly in small amplitude limit. 
Similar to one dimensional (1D) drift waves, the 2D nonlinear drift wave equation predicts the existence of 2D dip solitons. The HM equation contains only one nonlinear term that is the ion vorticity term which can be smaller than nonlinear electron density term  and that appears naturally due to the presence of Boltzmann electrons. If the wave amplitude is assumed to be slowly varying function of space, then the electron nonlinear term may not be smaller than ion vorticity term. 

In this work, we show that 1D and 2D electrostatic density cavities possessing widths in perpendicular direction to ${\bf B}_0$ can be created in the inhomogeneous upper ionosphere by nonlinear drift waves, if a simpler and appropriate theoretical model is used for the study of these waves.

	\section{Nonlinear Drift Waves}
	To study drift waves, we assume the external magnetic field to be along z-axis i.e. ${\bf B}_0= B_0 \hat{z}$ and density gradient to be along negative x-axis.
	Recently \parencite{saleem2026exploring}, the ion velocity ${\bf v}_i$ has been normalized with the electron diamagnetic drift velocity $v_{De}$ in the derivation of general nonlinear equation. Using following normalizations, $n_i^{\prime}=\frac{n_i}{n_0}$, ${\bf v}_i^{\prime}= \frac{{\bf v}_i}{v_{De}}$, $y^{\prime}=\frac{y}{\rho_s}$, $x^{\prime}=\frac{x}{\rho_s}$, $\nabla^{\prime}= \rho_s \nabla$, and $t^{\prime}=\frac{c_s}{L_n}t$ where $L_n=\frac{1}{\kappa_n}$, $\frac{c_s}{L_n}=\frac{v_{De}}{\rho_s}$, $v_{De}= D_e \kappa_n$ is the electrons diamagnetic drift velocity and $\kappa_n=\mid \frac{1}{n_0} \frac{dn_0}{dx}  \mid$, the nonlinear equation for drift waves has been expressed as,
	\begin{equation}
		\partial_t (\frac{\tilde{n_i}}{n_{i0}}- \nabla_{\perp}^2 \Phi) + (\nabla_{\perp} \Phi \times \hat{z}. \hat{x}) 
    \end{equation}
    $$
        + 
        \frac{L_n}{\rho_s} (\nabla_{\perp} \Phi \times \hat{z} \cdot \nabla_{\perp}) \nabla_{\perp}^2 \Phi =0
	$$
	where $\Phi=\frac{e \varphi}{T_e}$ and the superscript prime (${\prime}$) has been dropped for simplicity.
	This equation can be written in single variable $\Phi$ as follows,
	\begin{equation}
		\partial_t (\Phi + \frac{1}{2} \Phi^2- \nabla_{\perp}^2 \Phi) + (\nabla_{\perp} \Phi \times \hat{z}. \hat{x})
	\end{equation}
	
	$$
	+ \frac{L_n}{\rho_s} (\nabla_{\perp} \Phi \times \hat{z} \cdot \nabla_{\perp}) \nabla_{\perp}^2 \Phi =0
	$$
	Equation (2) is the same as Eq. (13) of Ref. \parencite{saleem2026exploring}. In physical units, it can be expressed as,
	\begin{equation}
		\partial_t (\Phi + \frac{1}{2} \Phi^2- \rho_s^2 \nabla_{\perp}^2 \Phi) + v_{De} (\nabla_{\perp} \Phi \times \hat{z} \cdot \hat{x})
	\end{equation}
	
	$$
	+ \Omega_i \rho_s^4 (\nabla_{\perp} \Phi \times \hat{z} \cdot \nabla_{\perp}) \nabla_{\perp}^2 \Phi=0
	$$
	where $\rho_s=\frac{c_s}{\Omega_i}$. In equation (3), only $\Phi$ is normalized.
	The linear dispersion relation for drift waves can be expressed as,
	\begin{equation}
		\omega= \frac{v_{De} k_y}{(1+\rho_s^2 k^2)}=\frac{\omega^{\ast}}{(1+\rho_s^2 k_y^2)}
	\end{equation}
	where $\omega^{\ast}={\bf v}_{De} \cdot {\bf k}_y$ is the drift wave frequency, ${\bf v}_{De}=v_{De} \hat{y}= \frac{c T_e}{e B_0} \nabla \ln n_0 \times \hat{z} =(\frac{cT_e}{eB_0} \kappa_n) \hat{y}$, and $\nabla n_0 =- \hat{x} \mid  \frac{dn_0}{dx}  \mid$. The wave propagates predominantly along y-direction perpendicular to both $\nabla n_0$ and ${\bf B}_0$.
	
	\section{Drift Solitons}
	If nonlinear ion vorticity is ignored, the Eq. (3) becomes,
	\begin{equation}
		\partial_t (\Phi + \frac{1}{2} \Phi^2- \rho_s^2 \nabla_{\perp}^2 \Phi) + v_{De} (\nabla_{\perp} \Phi \times \hat{z} \cdot \hat{x}) =0
	\end{equation}
	To obtain solitary waves, we define stretched coordinates
	$(\eta, \xi, \tau)$ for $\epsilon \ll1$ as follows,
	\begin{equation}
		\xi=\epsilon^{1/2} (y - U_0t); \eta = \epsilon^{1/2} x
	\end{equation}
	and
	\begin{equation}
		\tau = \epsilon^{3/2} t
	\end{equation}
	which give $\partial_y =\epsilon^{1/2} \partial_{\xi}$, $\partial_x = \epsilon^{1/2} \partial_{\eta}$ and $\partial_t = (- \epsilon^{1/2} U \partial_{\xi} + \epsilon^{3/2} \partial_{\tau})$ where the wave phase speed $U_0$ is normalized with $v_{De}$ and hence $U=\frac{U_0}{v_{De}}=1$.
	In order $\epsilon^{5/2}$, Eq. (5) takes the form,
	\begin{equation}
		[ \partial_{\tau} \Phi  -  v_{De} (\Phi \partial_{\xi} \Phi)  +
		v_{De} \rho_s^4  \partial_{xi} (\partial_{\eta}^2 + \partial_{\xi}^2) \Phi\simeq 0
	\end{equation}
	This equation is the limiting form of Zakharov-Kuznetsov (ZK) equation in $(\eta, \xi, \tau)$ coordinates. To obtain solitary wave solution of Eq. (8), we define a coordinate $\chi$ such that,
	\begin{equation}
		\chi=\xi + \lambda \eta -M_0 \tau
	\end{equation}
	where $M_0$ is velocity of the nonlinear wave and $\lambda < 1$ because the wave propagates along y-axis predominantly. Equation (7) in $\chi$ coordinate can be expressed in the following form,
	\begin{equation}
		- M_0 d_{\chi} \Phi - v_{De} \Phi d_{\chi} \Phi + v_{De} \rho_s^2 (1+\lambda^2) d_{\chi}^2 \Phi =0
	\end{equation}
	The above equation is written in simpler form as,
	\begin{equation}
		- d_{\chi} \Phi - A \Phi d_{\chi} \Phi + B d_{\chi}^3 \Phi =0
	\end{equation}
	where $A=\frac{1}{M}$, $B= \rho_s^2 (\frac{1 + \lambda^2}{M})$ and $M= \frac{M_0}{v_{De}}$ is the Mach number.
	Solution of this equation is,
	\begin{equation}\label{soliton profile}
		\Phi(\chi) = - \Phi_m [Sech^2 (\frac{\chi}{W})]
	\end{equation}
	where $\Phi_m= 3 M$ is the maximum amplitude and $W=2 \sqrt B$ the width of solitary pulse. The pulse will be 1D if $\lambda =0$, otherwise it will be 2D as shown in Ref. \parencite{saleem2026exploring}.
	
	\vspace{0.5 cm}
	
	\section{Density Cavities in Ionosphere}
	In this section we apply the results to the upper ionospheric plasma at altitude of $1700$ $km$ where Freja satellite has observed the density cavities. The size along the field has not been determined in Ref. \parencite{dovner1994freja}. On the other hand, for a particular observation the width of such a structure in direction perpendicular to ${\bf B}_0$ has been  quoted to be $15$ $m$ and magnitude of potential in an observation has been mentioned to be about $60$ $mV$. These observed values provide a clear guess of the order of magnitude of such electrostatic structures in upper ionosphere. It should be noted that $B \propto \rho_s^2$ and hence $W \propto \rho_s$ which is a few meters in this region. Therefore, the size of the density cavities should turn out to be a few or several meters depending upon the values of $M$ and $\lambda$.
	
	\subsection{1D Density Cavities}
	\begin{table}[htbp!]
		\centering
		\label{tab:plasma_parameters_M1}
		\renewcommand{\arraystretch}{1.2}
        \begin{tabular}{cccccc}
			\hline
			$M$  & $\phi_m$ (mV) & $W$ (m) \\
			\hline
			0.06  & 17.9999 & 40.0621 \\
			0.08  & 23.9999 & 34.6948 \\
			0.10  & 29.9998 & 31.0319 \\
			\hline
		\end{tabular}
		\caption{Maximum amplitude and width of solitary waves corresponding to different
			values of Mach number $M$ for fixed $B_0=0.26$ G and $\lambda=0$.}
	\end{table}
	
	\begin{figure}[htbp!]
		\centering
		\includegraphics[width=0.75\linewidth]{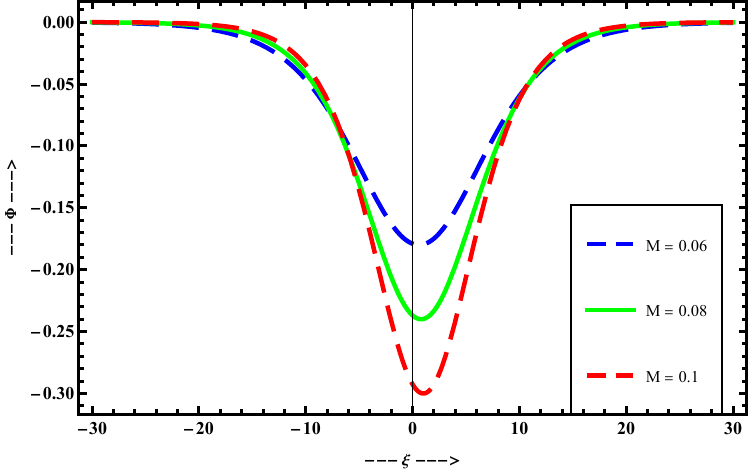}
		\caption{The 1D soliton profiles ($\lambda=0$) for $B_0=0.26$ G and three different values of the Mach number $M$ evaluated at $\tau_0=10$ s using Eq. \eqref{soliton profile}} 
		\label{fig:1(a)}
	\end{figure}
	\begin{figure}[htbp!]
		\centering
		\includegraphics[width=0.75\linewidth]{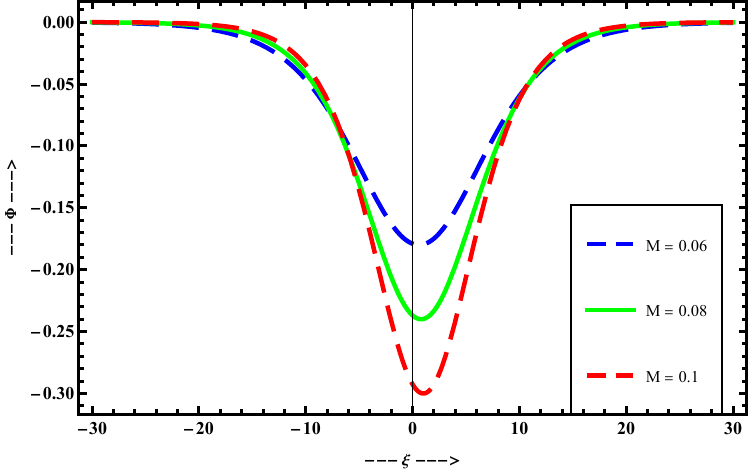}
		\caption{The 1D soliton profiles ($\lambda=0$) for  $B_0=0.4$ G and three different values of Mach number $M$ evaluated at $\tau_0=10$ s using Eq. \eqref{soliton profile}}
		\label{fig:1(b)}
	\end{figure}
	
	Let us just estimate the size of 1D structure assuming $\lambda=0$ in the solution given by Eq. (12).
	We use the typical parameters of ionosphere at $1700$ $km$ as $T_e= (0.1)$ $ev$ and $B_0=(0.26)$ $G$ whereas the plasma is considered to be the pure oxygen plasma. We find $\rho_s \simeq 3$ $m$, $W \simeq 20$ $m$ and $\mid \varphi \mid \simeq 30$  $mV$, if we choose $ M=(0.1) $. The electric field estimate turns out to be $\mid E \mid \simeq (1.5)$ $mV/m$. Figure (1) illustrates the form of negative electrostatic potential with $\Phi_m = 0.3$.
	
	
	\subsection{2D Density Cavities in Ionosphere}
	
	\begin{table}[htbp]
		\centering
		
		\label{tab:plasma_parameters_M2}
		\renewcommand{\arraystretch}{1.2}
		\begin{tabular}{cccccc}
			\hline
			$M$ & $\phi_m$ (mV) & $W$ (m) \\
			\hline
			0.06 &  17.9999 & 40.2619 \\
			0.08 &  23.9999 & 34.8678 \\
			0.10 &  29.9998 & 31.1867 \\
			\hline
		\end{tabular}
		\caption{Estimates for the magnitudes of potential and width of solitary structures corresponding to different
			values of Mach number $M$, for $B_0=0.26$ G and  $\lambda=0.1$.}
	\end{table}
	
		
			
	\begin{table}[htbp]
		\centering
		
		\label{tab:plasma_parameters}
		\renewcommand{\arraystretch}{1.2}
		\begin{tabular}{cccccc}
			\hline
			$\lambda$ & $\phi_m$ (mV) & $W$ (m)\\
			\hline
			0.1 & 29.9998 & 31.1867 \\
			0.5 & 29.9998 & 34.6948 \\
			1.0 & 29.9998 & 43.8858 \\
			\hline
		\end{tabular}
		\caption{Change in width corresponding to different values of $\lambda$ for $B_0=0.26$ G and Mach number $M=0.1$.}
	\end{table}
	\begin{figure}[htbp!]
		\centering
		\includegraphics[width=\linewidth]{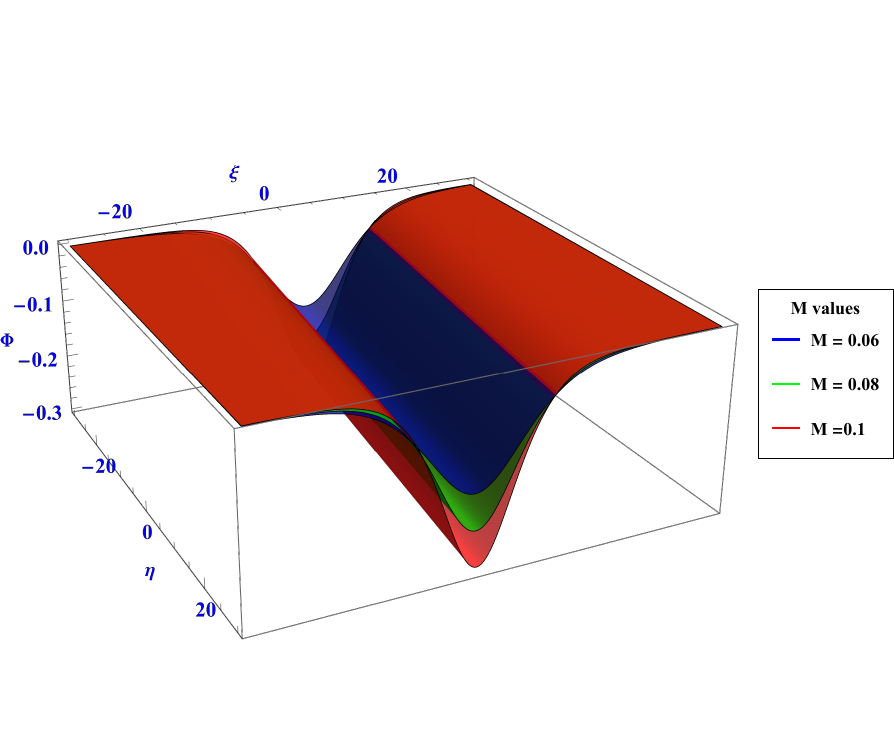}
		\caption{2D Soliton profile  for $B_0=0.26$ G and different values of the Mach number $M$ evaluated for $\lambda=0.1$ at $\tau_0=10$ s \eqref{soliton profile}}
		\label{fig:2(a)}
	\end{figure}
	\begin{figure}[htbp!]
		\centering
		\includegraphics[width=\linewidth]{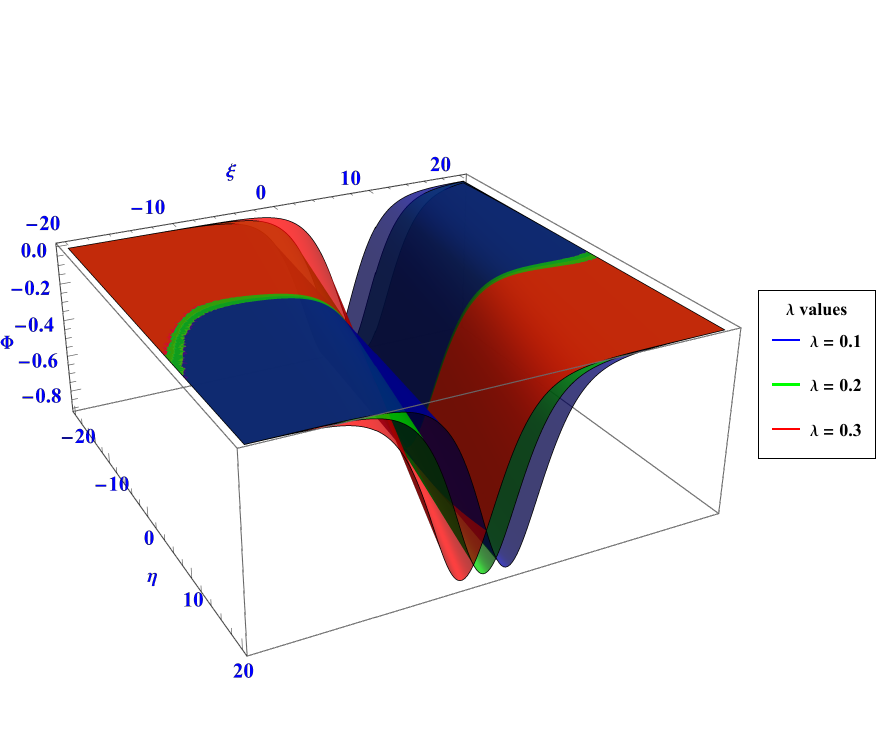}
		\caption{2D soliton profile for $B_0=0.26$ G and different values of $\lambda$ evaluated for $M=0.3$ at $\tau_0=10$ s using Eq. \eqref{soliton profile}}
		\label{fig:2(b)}
	\end{figure}
	If we choose $\lambda \neq 0$, then the density cavity will have a 2D form in $\xi\eta$-plane. The amplitude is independent of $\lambda$ and width of the structure changes corresponding to different values of this parameter.

	\newpage
	
	\section{Discussion}
A creation mechanism of electrostatic density cavities by nonlinear drift waves in the upper ionosphere has been proposed. The linear drift waves are considered usually to propagate in direction perpendicular to both density gradient $\nabla n_0$ and ambient magnetic field ${\bf B}_0$. The component of wave vector parallel to magnetic field $k_z$ must not be exactly zero for the existence of drift waves because inertia-less electrons can follow Boltzmann distribution only if the component of wave vector parallel to the magnetic field is non-zero $k_z \neq 0$. On the other hand, if the contribution of ion parallel motion is considered, then the ion acoustic waves couple with the drift waves and it becomes difficult to differentiate the role of these two low frequency modes.

Several decades ago, Hasegawa and Mima \parencite{hasegawa1978pseudo} derived a nonlinear equation for pure drift waves by ignoring the contribution of ion parallel motion assuming $k_z \simeq 0$. The amplitude of nonlinear drift waves was assumed to be a function of both $(x,y)$ coordinates for ${\bf B}_0= B_0 \hat{z}$. The well-known Hasegawa-Mima (HM) equation explains the drift wave turbulence observed in tokamak plasma. In addition to it, this equation also admits a stationary solution in the form of dipolar vortices. This equation was derived assuming $\rho_s k_{\perp} \simeq 1$ and $\frac{\rho_s}{L_n} \simeq \Phi \simeq \epsilon$ where $\epsilon \ll 1$ was a smallness parameter. In this limit, the nonlinear electron density term becomes of higher order and is ignored. But the electron density nonlinear term is a natural consequence of electrons Boltzmann distribution and should be taken into account in the study of drift waves as has been discussed in Ref. \parencite{saleem2026exploring}. When both the electron density nonlinear term (scalar nonlinearity) and the nonlinear ion vorticity term (vector nonlinearity) are taken into account, then the nonlinear drift wave equation becomes complicated and its analytical solution has not been found yet.

However, this equation reduces to the limiting case of ZK equation which admits the solution in the form of 2D solitons. Interestingly, the ZK form of equation for drift waves admits the dip soliton solution which are often observed in space plasmas. If only 1D case is considered, then ZK equation reduces to KdV equation and yields 1D solitons. Thus the study of pure drift waves seems to be important to explain mechanism for the generation of density cavities.

Ion acoustic waves (IAWs) are also low frequency waves and electrons are considered to be inertia-less in the electrostatic field of these waves. But IAWs produce hump solitons and cannot explain the formation mechanism of density cavities if electrons are assumed to follow Boltzmann distribution. Therefore, Cairn's et al. \parencite{Cairns1995} considered the electrons to be non-thermal to explain the creation mechanism of density cavities along field lines. They considered propagation of IAWs along field lines and showed that within narrow range of parameters, the density cavities along ${\bf B}_0$ can be produced by IAWs if electrons follow the non-thermal Cairn's distribution. 

But Feja satellite has observed the density cavities which also have widths in direction perpendicular to the external magnetic field and presence of Boltzmann electrons has also been detected \parencite{dovner1994freja}. Freja satellite observed two types of electrostatic density cavities in upper ionospheric plasma, one is associated with the lower hybrid waves and the other type exists in the presence of Boltzmann electrons which gives a clue for the low frequency wave activity in this region. The upper ionosphere is inhomogeneous and magnetized and hence the present theoretical model is applicable and explains the mechanism for the creation of density cavities with widths in direction perpendicular to the geomagnetic field. However, the density depletion along field lines needs to be studied by taking into account the role of ion parallel motion as well which is out of scope of the present work.

	\bigskip
	\printbibliography

@article{dovner1994freja,
  title={Freja multiprobe observations of electrostatic solitary structures},
  author={Dovner, PO and Eriksson, AI and Bostr{\"o}m, R and Holback, B},
  journal={Geophysical Research Letters},
  volume={21},
  number={17},
  pages={1827--1830},
  year={1994},
  publisher={Wiley Online Library}
}

@article{Wahlund1994a,
  title={On ion acoustic turbulence and the nonlinear evolution of kinetic Alfv{\'e}n waves in aurora},
  author={Wahlund, J-E and Louarn, P and Chust, T and De Feraudy, H and Roux, A and Holback, B and Dovner, P-O and Holmgren, G},
  journal={Geophysical Research Letters},
  volume={21},
  number={17},
  pages={1831--1834},
  year={1994},
  publisher={Wiley Online Library}
}

@article{Wahlund1994b,
  title={Observations of ion acoustic fluctuations in the auroral topside ionosphere by the FREJA S/C},
  author={Wahlund, J-E and Louarn, P and Chust, T and De Feraudy, H and Roux, A and Holback, B and Cabrit, B and Eriksson, AI and Kintner, PM and Kelley, MC and others},
  journal={Geophysical Research Letters},
  volume={21},
  number={17},
  pages={1835--1838},
  year={1994},
  publisher={Wiley Online Library}
}

@article{eriksson1994freja,
  title={Freja observatons of correlated small-scale density depletions and enhanced lower hybrid waves},
  author={Eriksson, AI and Holback, B and Dovner, PO and Bostr{\"o}m, R and Holmgren, G and Andr{\'e}, M and Eliasson, L and Kintner, PM},
  journal={Geophysical Research Letters},
  volume={21},
  number={17},
  pages={1843--1846},
  year={1994},
  publisher={Wiley Online Library}
}

@article{bostrom1988characteristics,
  title={Characteristics of solitary waves and weak double layers in the magnetospheric plasma},
  author={Bostr{\"o}m, Rolf and Gustafsson, Georg and Holback, Bengt and Holmgren, Gunnar and Koskinen, Hannu and Kintner, Paul},
  journal={Physical Review Letters},
  volume={61},
  number={1},
  pages={82},
  year={1988},
  publisher={APS}
}

@article{Cairns1995,
  title={Electrostatic solitary structures in non-thermal plasmas},
  author={Cairns, RA and Mamum, AA and Bingham, R and Bostr{\"o}m, R and Dendy, RO and Nairn, CMC and Shukla, PK},
  journal={Geophysical Research Letters},
  volume={22},
  number={20},
  pages={2709--2712},
  year={1995},
  publisher={Wiley Online Library}
}

@misc{sagdeev1966reviews,
  title={Reviews of plasma physics},
  author={Sagdeev, RZ and Leontovich, MA},
  year={1966},
  publisher={Consultants Bureau New York}
}

@article{washimi1966,
  title={Propagation of ion-acoustic solitary waves of small amplitude},
  author={Washimi, Haruichi and Taniuti, Tosiya},
  journal={Physical Review Letters},
  volume={17},
  number={19},
  pages={996},
  year={1966},
  publisher={APS}
}

@article{yu1980fully,
  title={Fully nonlinear ion-acoustic solitary waves in a magnetized plasma},
  author={Yu, MY and Shukla, PK and Bujarbarua, S},
  journal={The Physics of Fluids},
  volume={23},
  number={10},
  pages={2146--2147},
  year={1980},
  publisher={AIP Publishing}
}

@article{steffy2017existence,
  title={Existence domain of the compressive ion acoustic super solitary wave in a two electron temperature warm multi-ion plasma},
  author={Steffy, SV and Ghosh, SS},
  journal={Physics of Plasmas},
  volume={24},
  number={10},
  year={2017},
  publisher={AIP Publishing}
}

@article{saleem2020theoretical,
  title={Theoretical models for unstable IAWs and nonlinear structures in the upper ionosphere},
  author={Saleem, H and Shan, S Ali},
  journal={Reviews of Modern Plasma Physics},
  volume={4},
  number={1},
  pages={3},
  year={2020},
  publisher={Springer}
}

@article{lakhina2021electrostatic,
  title={Electrostatic solitary structures in space plasmas: Soliton perspective},
  author={Lakhina, Gurbax Singh and Singh, Satyavir and Rubia, Rajith and Devanandhan, Selvaraj},
  journal={Plasma},
  volume={4},
  number={4},
  pages={681--731},
  year={2021},
  publisher={MDPI}
}

@article{varghese2022morphology,
  title={On the morphology of electrostatic solitary waves in the Earth’s aurora},
  author={Varghese, Steffy Sara and Singh, Kuldeep and Kourakis, Ioannis},
  journal={Scientific Reports},
  volume={12},
  number={1},
  pages={18204},
  year={2022},
  publisher={Nature Publishing Group UK London}
}

@article{johnson1995nonlinear,
  title={Nonlinear vortex structures with diverging electric fields and their relation to the black aurora},
  author={Johnson, Jay R and Chang, Tom},
  journal={Geophysical Research Letters},
  volume={22},
  number={12},
  pages={1481--1484},
  year={1995},
  publisher={Wiley Online Library}
}

@article{shukla1995vortex,
  title={Vortex streets driven by sheared flow and applications to black aurora},
  author={Shukla, PK and Birk, GT and Bingham, R},
  journal={Geophysical Research Letters},
  volume={22},
  number={6},
  pages={671--674},
  year={1995},
  publisher={Wiley Online Library}
}

@article{saleem2026exploring,
  title={Exploring nonlinear drift waves: Limiting cases and dynamics},
  author={Saleem, H},
  journal={AIP Advances},
  volume={16},
  number={1},
  year={2026},
  publisher={AIP Publishing}
}

@Article{Saleem2024,
  author    = {Saleem, H.},
  journal   = {Physics of Plasmas},
  title     = {Drift wave KdV equation creates dip solitons},
  year      = {2024},
  issn      = {1089-7674},
  month     = nov,
  number    = {11},
  volume    = {31},
pages={112102},
  %doi       = {10.1063/5.0230434},
  publisher = {AIP Publishing},
}

@article{nozaki1974propagation,
  title={Propagation of drift waves of small but finite amplitude},
  author={Nozaki, Kazuhiro and Taniuti, Tosiya},
  journal={Journal of the Physical Society of Japan},
  volume={36},
  number={1},
  pages={267--270},
  year={1974},
  publisher={The Physical Society of Japan}
}

@article{gell1977drift,
  title={Drift solitons and their two-dimensional stability},
  author={Gell, Y},
  journal={Physical Review A},
  volume={16},
  number={1},
  pages={402},
  year={1977},
  publisher={APS}
}

@article{hasegawa1978pseudo,
  title={Pseudo-three-dimensional turbulence in magnetized nonuniform plasma},
  author={Hasegawa, Akira and Mima, Kunioki},
  journal={The Physics of Fluids},
  volume={21},
  number={1},
  pages={87--92},
  year={1978},
  publisher={American Institute of Physics}
}
	
\end{document}